# The Burst Alert Telescope (BAT) on the *Swift* MIDEX mission

Scott D. Barthelmy[1], Louis M. Barbier[1], Jay R. Cummings[1,2], Ed E. Fenimore[3], Neil Gehrels[1], Derek Hullinger[1,4], Hans A. Krimm[1,5], Craig B. Markwardt[1,4], David M. Palmer[3], Ann Parsons[1], Goro Sato[6,7], Masaya Suzuki[6,7], Tadayuki Takahashi[6,7], Makota Tashiro[6,7], Jack Tueller[1]

NASA Goddard Space Flight Center
Code 661, Greenbelt, MD, 20771

## ABSTRACT

The Burst Alert Telescope (BAT) is one of 3 instruments on the *Swift* MIDEX spacecraft to study gamma-ray bursts (GRBs). The BAT first detects the GRB and localizes the burst direction to an accuracy of 1-4 arcmin within 20 sec after the start of the event. The GRB trigger initiates an autonomous spacecraft slew to point the two narrow field-of-view (FOV) instruments at the burst location within 20-70 sec so to make follow-up x-ray and optical observations. The BAT is a wide-FOV, coded-aperture instrument with a CdZnTe detector plane. The detector plane is composed of 32,768 pieces of CdZnTe (4x4x2mm), and the coded-aperture mask is composed of ~52,000 pieces of lead (5x5x1mm) with a 1-m separation between mask and detector plane. The BAT operates over the 15-150 keV energy range with ~7 keV resolution, a sensitivity of $\sim 10^{-8}$ erg $s^{-1}$ $cm^{-2}$, and a 1.4 sr (half-coded) FOV. We expect to detect >100 GRBs/yr for a 2-year mission. The BAT also performs an all-sky hard x-ray survey with a sensitivity of ~2 mCrab (systematic limit) and it serves as a hard x-ray transient monitor.

## 1. INTRODUCTION

### 1.1 *Swift* Mission Overview

From the discoveries by *BeppoSAX*[1] and the subsequent ground-based follow-ups[2,3], we now know that gamma-ray bursts (GRB) are cosmological (red-shifts range from 0.0085 to ~4.5). As of September 2004, there are 34 GRBs and two X-ray flashes with firm redshift measurements. *Swift* will add greatly to this number, and by studying the frequency histogram distribution with redshift, it will become possible to distinguish between a simple model in which GRBs simply trace the comoving volume of the Universe, and more detailed models that trace, for example, star formation. Historically, the difficulties involved with operations and scheduling constrain the time delay between the burst and when the small error-box positions are available, and therefore when the follow-up observations can begin, to be typically in the range of 3 to 8 hours. These delays hold for all but two of the ~100 GRBs that have follow-up observations -- GRB990123 was observed in the optical during the burst itself[4]. These time delays are rather limiting to the study of GRB engines. To constrain GRB theories in a meaningful way, a much shorter response time is required.

*Swift* will eliminate these delays and fill in the hole in the afterglow observations by using in a single spacecraft

1NASA Goddard Space Flight Center, Laboratory for High Energy Astrophysics, Greenbelt, MD 20771
2National Research Council, 2101 Constitution Avenue, NW, Washington, DC 20418
3Los Alamos National Laboratory, P.O. Box 1663, Los Alamos, NM 87545
4Department of Astronomy, University of Maryland, College Park, MD 20742-2421
5Universities Space Research Association, 10227 Wincopin Circle, Suite 212, Columbia, MD 21044
6Institute of Space and Astrononautical Science, 3-1-1, Yoshinodai, Sagamihara, 229-8510 Kanagawa, Japan
7Department of Physics, University of Tokyo, Hongo 7-3-1, Bunkyo-ku, 113 Tokyo, Japan

instrument to detect the burst, provide a position within a several seconds, incorporate on-board autonomy to execute a spacecraft slew without ground-based intervention, and point x-ray and UV/optical telescopes at the burst position to make follow-up observations within as little as 20-70 seconds after the start of the burst. For ~10% of the bursts, the x-ray and UV/optical follow-up observations will begin while the GRB is still bursting in the gamma-ray bandpass. UV observations are not possible from ground-based instruments. It is this rapid-response, plus panchromatic approach, that will provide the next step in our understanding of GRBs. The *Swift* compliment of instruments also provides a powerful observatory for the follow-up of other hard x-ray transients detected by BAT.

### 1.1.1 The *Swift* Science

There are several key questions which the *Swift* mission will address: What are the sites and nature of the GRB progenitors? Are there multiple classes of GRBs? What are the local environments around the progenitors? What can GRBs tells us about the early universe? For a more complete discussion of these questions and a description of the science objectives by *Swift,* see Gehrels[5].

### 1.1.2 The *Swift* Instruments

The Burst Alert Telescope (BAT) is a highly sensitive, large FOV, coded-aperture telescope designed to monitor a large fraction of the sky for the occurrences of GRBs. The BAT provides the burst trigger and the 1-4 arcmin accurate position, that is then used to slew the spacecraft to point the two narrow-FOV instruments (the X-ray telescope – XRT, and the ultraviolet and optical telescope - UVOT) for follow-up observations. While observing bursts, BAT simultaneously and automatically accumulates an all-sky hard x-ray survey. The BAT consists of a 5200 $cm^2$ array of 4x4 $mm^2$ CdZnTe elements located 1 meter behind a 2.7 $m^2$ coded-aperture mask of 5x5 $mm^2$ elements, with a point spread function (PSF) of 17 arcmin. The BAT coded-aperture mask, and hence its FOV, is limited by the Delta rocket faring. The BAT instrument was designed and built at Goddard Space Flight Center. The details of BAT are described in section 2.

The XRT is a sensitive, autonomous x-ray CCD imaging spectrometer at the focal plane of a 3.5 m focal length, nested grazing incidence Wolter type I telescope. It covers an energy range of 0.2 - 10 keV with a 23x23 square arcmin FOV and 3 arcsec positioning capability. The details of XRT are described in Burrows[6] and Gehrels[5]. The UVOT is an optical and UV 30-cm aperture Ritchey-Chretien telescope. It has 6 bandpass filters operating over a range of 170-650 nm, plus two grisms. The FOV is 17x17 square arcmin with a sensitivity of $B$=24 magnitude in a 1000-sec exposure. It can centroid sources to an accuracy of 0.3 arcsec. The details of the UVOT are described in Roming[7] and Gehrels[5]. The XRT and UVOT are co-aligned and pointed near the center of the BAT FOV.

### 1.1.3 The *Swift* Spacecraft

The spacecraft bus was built by Spectrum Astro (Gilbert, AZ). It is shown in Figure 1. It is a 3-axis stabilized platform. The slew rate has been enhanced beyond typical platforms because of the rapid slewing requirements for the narrow-FOV instrument (NFI) follow-up on the BAT burst positions. The platform can slew from 0 to 50° in 20-70 sec. Also, because of the rapid follow-up requirement, the Attitude Control System (ACS) has full on-board autonomy for conducting a slew maneuver. No ground mission operations decisions or commanding are needed for the spacecraft to calculate and initiate a slew maneuver. The on-board autonomy checks all spacecraft orientation constraints (sun angle, moon angle, earth limb angle, and ram vector). If the calculated slew to the new BAT burst position is determined not to violate any of the constraints, then a slew is performed. This will happen ~90% of the time. Once a new target has been acquired, the pointing stability is better than 0.1 arcsec, which is more precise than is needed for BAT and is driven by the requirements for the two NFIs.

The spacecraft has a solid-state data recorder with enough capacity for ~3 days of operations. The BAT survey data dominates the spacecraft telemetry rate. Typically, the data will be down-linked ~7-10 times per day to the Malindi ground station. Commands will also be uplinked through Malindi. Additionally, the spacecraft can transmit small quantities of data in real-time directly to TDRSS using the Multiple Access (MA) capability. The MA capability is used to get the burst position, lightcurve, spectra, and x-ray and optical finding-chart images to the ground within seconds to minutes after the burst.

### 1.1.4 *Swift* In-Flight Operations

The *Swift* spacecraft carries out the following observation scheme. The two NFIs carry out a series of follow-up observations on a previously detected burst. The types of observations and their durations are described in Burrows[6] and Roming[7]. While the NFIs are observing, the BAT instrument is staring at a region of sky waiting for a new burst to happen (and also accumulating the survey mode data). In general there will be several bursts for which the NFIs are conducting a series of follow-up observations. At any given instant, which prior burst is being observed and in what manner is determined by an observing table uploaded from the Mission Operations Center (MOC). Prior observations, Earth occultations, etc. determine the content of the uploaded observing plan. This observing table contains a week's worth of observations. The decision to interrupt the current observation with a slew to a newly detected BAT burst position is determined by the Figure Of Merit algorithm (see section 2.3.3).

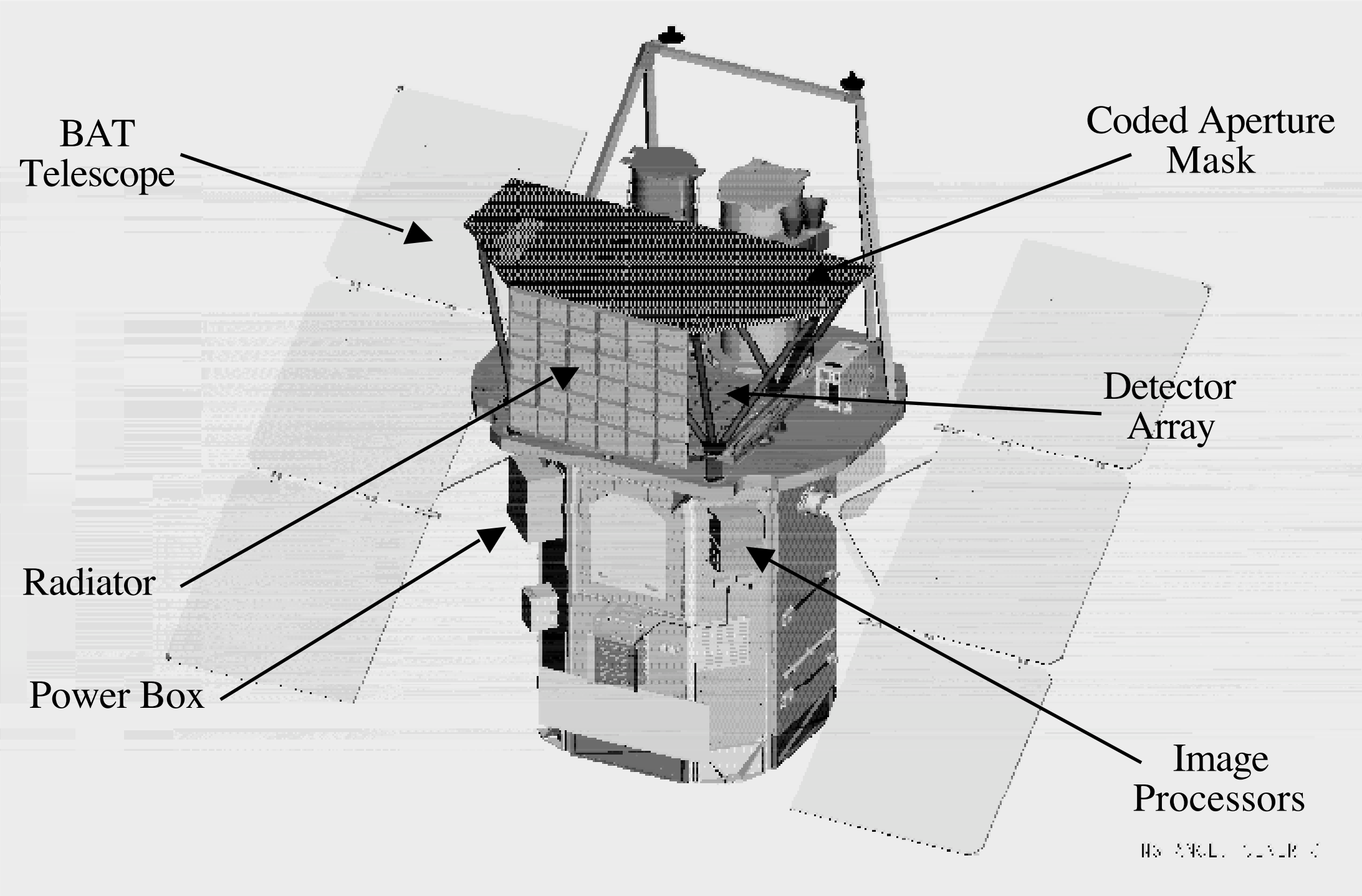

**Figure 1.** The *Swift* spacecraft showing the BAT (the D-shaped coded-aperture mask), the XRT and UVOT instruments (the two tubes in the back), and the solar panels. BAT is composed of the Detector Array, the Coded Mask, the thermal Radiator, the 2 Image Processors, and the Power Control box. The Graded-Z Fringe Shield side walls have been removed to show the Detector Array and the Struts that support the Mask.

### 1.1.5 Mission Profile

Table 1 contains the basic mission profile parameters. The orbital radius of 600 km was selected to keep the instrument background low from the Van Allen belts, plus be high enough to guarantee a longer than 2-year orbit lifetime. The worst-case solar activity scenarios predict a minimum orbit lifetime of 5 years. The Delta 7320 launch vehicle provides a further 280 kg mass margin, which will be used in a dog-leg maneuver during launch to decrease the orbit inclination from the 28.5° to something below 22°. The lower inclination minimizes the amount of time in the South Atlantic Anomaly (SAA) and increases the BAT's orbital averaged livetime – allowing ~10% more bursts to be detected.

**TABLE 1.** *Swift* Mission Launch and Orbital Characteristics.

| **Mission Parameter** | **Value** |
|---|---|
| Launch Date | November 2004 |
| Launch Vehicle | Delta 7320 |
| Orbit | ~600 km circular, <22° inclination |
| Total Mass | 1270 kg |
| Total Power | 1650 W |
| Mission life, Orbital lifetime | 2 yr; >5 yr |

# 2. THE BAT INSTRUMENT

The BAT instrument is shown in Figure 1. The Burst Alert Telescope (BAT) makes the initial detection of the gamma-ray burst (GRB), calculates a position for that burst, makes an on-board decision if the burst is worth an NFI follow-up observation, and sends that position to the spacecraft attitude control system, if it is worthy. It does all this within 12 sec of the initial trigger of the burst. To do this for a large number of bursts (>100 $yr^{-1}$), BAT has a large FOV (1.4 sr half-coded & 2.3 sr partially-coded). The only way to image such a large FOV is to use the coded-aperture technique. The following sections describe the details of the design, the function of the BAT instrument, and the data products that will be available to the world community.

## 2.1 Technical Description

The basic numbers describing the BAT instrument are listed in Table 2. The BAT instrument consists of a detector plane of 32,768 CZT detector elements and front-end electronics (section 2.1.1), a coded aperture mask located 1 m above the detector plane (section 2.1.2), a graded-Z fringe shield to reduce the instrumental background event rate and cosmic diffuse background (section 2.1.3), and a thermal radiator and control system to keep the detector plane at a constant temperature (section 2.1.4). The control of the BAT instrument is done through the Image Processor (section 2.1.5) and it also does the on-board event processing (burst trigger detection, burst location calculations, and burst figure-of-merit calculation (section 2.3.3)). While searching for bursts (section 2.3.1), BAT also accumulates a hard x-ray survey of the entire sky (section 2.3.2) over the course of the mission.

**TABLE 2.** *Swift*-BAT instrument parameters.

| **Parameter** | **Value** |
|---|---|
| Energy Range | 15-150 keV |
| Energy Resolution | ~7 keV |
| Aperture | Coded mask, random pattern, 50% open |
| Detecting Area | 5240 $cm^2$ |
| Detector Material | CdZnTe (CZT) |
| Detector Operation | Photon counting |
| Field of View | 1.4 sr (half-coded) |
| Detector Elements | 256 Modules of 128 elements/Module |
| Detector Element Size | 4.00 x 4.00 x 2.00 $mm^3$ |
| Coded Mask Cell Size | 5.00 x 5.00 x 1.00 $mm^3$ Pb tiles |
| Instrument Dimensions | 2.4m x 1.2m x 1.2 m |
| Telescope PSF | 17 arcmin |
| Source Position Accuracy | 1-4 arcmin |
| Sensitivity | ~$10^{-8}$ erg $s^{-1}$ $cm^{-2}$ |
| Number of bursts detected | >100 $yr^{-1}$ |

### 2.1.1 The Detector Plane

The BAT detector plane is composed of 32,768 pieces of 4.00 mm square CdZnTe (CZT) 2.00 mm thick. The CZT is supplied by eV Products Inc (Saxonburg, PA). For electronic control, event data handling, and fabrications reasons, these 32K pieces are grouped into two layers of hierarchical structure. Figure 2 shows this mechanical and electrical hierarchy. The hierarchy breaks down into the following definitions. A sub-array of 8x16 pieces of CZT are attached to a single analog signal processing ASIC (the XA1). Two of these sub-arrays with the ASIC control electronics constitutes a Detector Module (DM). Eight DMs are mounted in a mechanical structure called a Block. There are 16 Blocks mounted in the Detector Array Plane (DAP) in a 2-by-8 configuration. The mechanics of mounting the pieces of CZT yield a pixel pitch of 4.20 mm with gaps between the DMs and Blocks. The image reconstruction process (see section 2.3.1.2) easily handles these gaps, because the gaps have been restricted to be an integer multiple (2 or 3) of the basic pixel pitch. This division of the 32K pixels into 128 DMs and 16 Blocks, each with its own electronics, power supplies, control parameters, and communications links, provides a natural parallelism and redundancy against failures in the BAT instrument. This, plus the forgiving nature of the coded aperture technique, means that BAT can tolerate losses of individual pixels, individual DMs, and even whole Blocks without loosing the ability to detect bursts and determine locations. There is, of course, a loss in burst-detection and the survey sensitivities.

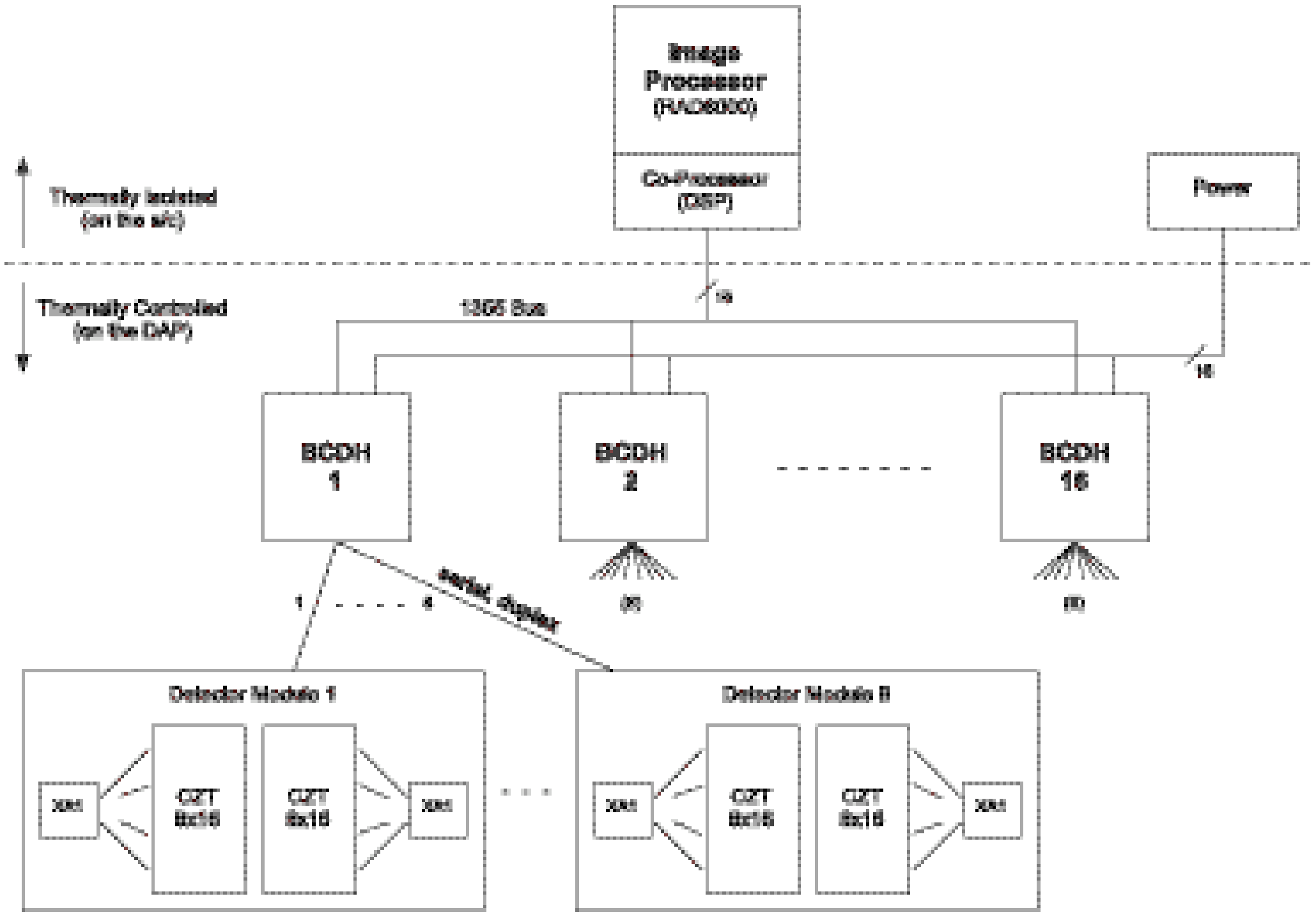

**Figure 2.** The BAT has the 32,768 pieces of CZT divided into the hierarchical structure shown here. This structure is implemented in both the mechanical grouping of the detector pixels and in the electronic control and processing of the events. It consists of 8x16 sub-arrays of CZT, two to a Detector Module, and 8 DMs in a Block, with 16 Blocks for the whole Detector Array.

#### 2.1.1.1 The Detector Module (DM)

The CZT pixel elements have planar electrodes and are biased typically to -200 volts. This bias voltage is commandable at the half-DM level (i.e. each sub-array of 128 pixels has a commandable HV supply of negative 0-300 volts). The front electronics on each DM are shown in Figure 3. The anode of each CZT piece is AC-coupled to an input on the XA1 ASIC along with a bleed resistor to ground for the CZT leakage current. The XA1 ASIC is made by IDE AS (Hovik, Norway). It has 128 channels of charge sensitive pre-amps, shaping amps, and discriminators. It is a self-triggering device. When supplied with a threshold control voltage, the ASIC will recognize an event on one of its 128 input channels, block out the remaining 127 channels, and present the pulse height of the event on its output for digitization. An on-board ADC unit digitizes the pulse height to a quantization of 0.5 keV. The XA1 has a linear range up to ~200 keV and extends up to ~500 keV with decreasing linearity. Because the XA1 also outputs the channel information for the event in analog form, the DM controller operates the analog multiplexer module (mux) to sequentially digitize the pulse height and address (segment and strip). Segment and strip addresses are combined by the controller into a single channel ID. The pulse height and channel ID are transmitted to the Block Controller and Data Handler (BCDH; see next section). This process takes 100 $\mu$sec. Individual channels can be logically disabled to handle noisy detectors. An in-house, custom, low-voltage differential signaling transceiver ASIC is used for the command and data links between the DMs and the BCDH. There is an electronic calibration pulser circuit within each DM. When commanded, it sequentially injects the specified number of charge pulses of the specified level into each of the 128 channels. These "calibration" events are flagged to be recognizable to the event processing routines in the Image Processor (section 2.1.5). This calibration pulser allows the offset, gain, and linearity of each of the 32K channels within the instrument to be monitored. In addition to the electronic calibration, there are two Am241 "tagged sources" which provide 60 keV photons for calibrating the absolute energy scale and detector efficiency of each CZT detector in-flight. The DM is capable of handling 20,000 events/sec which is sufficient for the expected brightest burst of the year of 700 events/sec/DM. Figure 4 shows an actual flight DM.

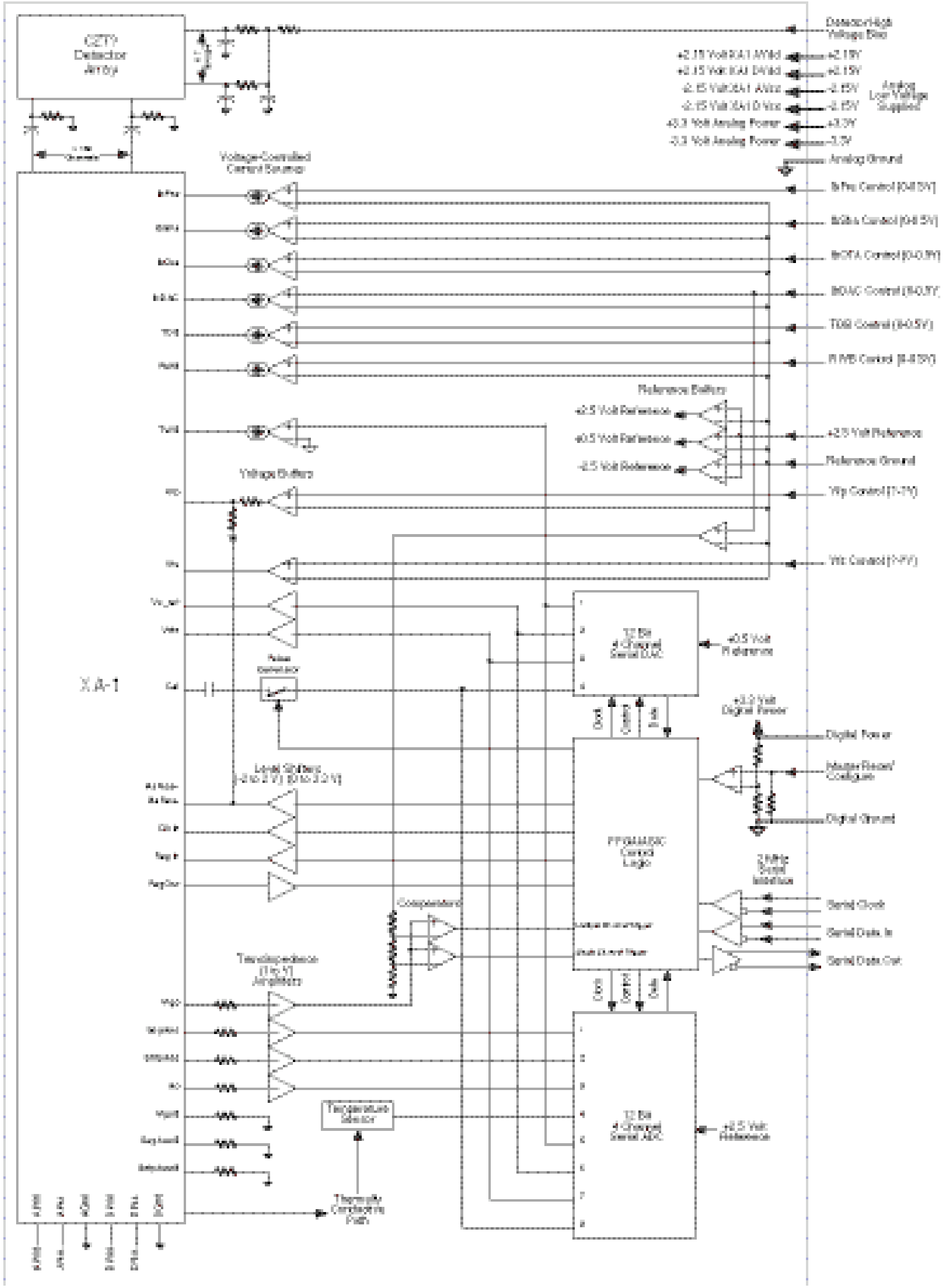

**Figure 3.** Block diagram of the front-end electronics for the BAT Detector Modules.

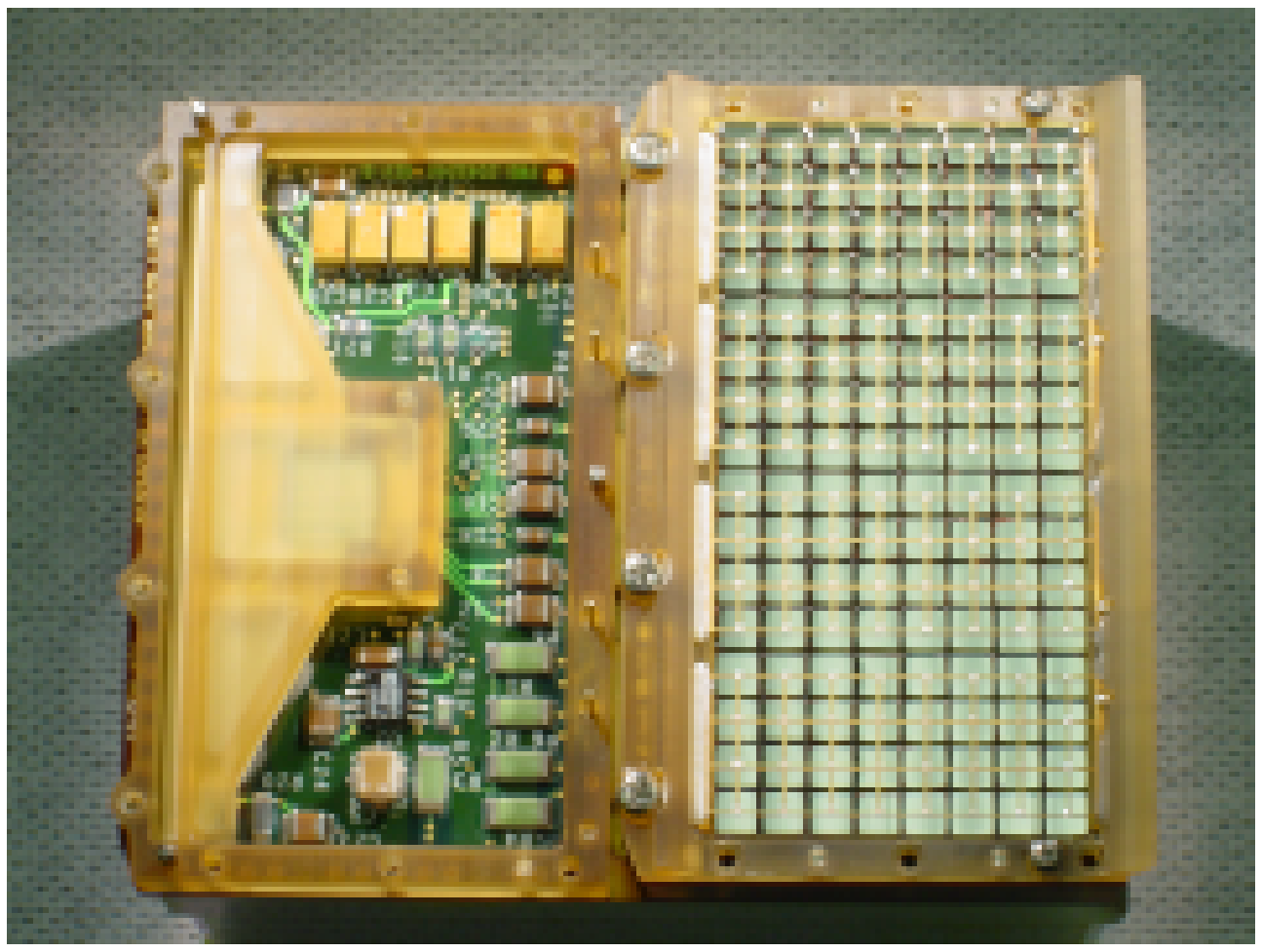

**Figure 4.** The BAT Detector Module (DM). Two sub-arrays of 8x16 pieces of CZT tile the top surface of the DM. The XA1 ASIC, mux, ADC, controller, and transceiver are located on Power Control Boxes (PCBs) stacked under the CZT layer. Eight of these DMs are packaged into a "Block" unit. The DM shown here has the aluminized-mylar EMI cover removed and the CZT layer removed on the left to show the XA1 ASIC under its Ultem protective cover.

#### 2.1.1.2 The Block and the Block Controller and Data Handler

The Block is the next level up in the hierarchy. The mechanical structure sits on the Detector Array Plate and holds 8 DMs, the Block Interface Card (BIC), the Block Voltage Regulator (BVR), and the Block Controller and Data Handler (BCDH). Figure 5 is a block diagram of these items (plus the Image Processor (IP), section 2.1.5). The BIC serves mostly as a mechanical and electrical connector routing interface between the DMs and the Block plus BCDH and BVR cards. The BVR provides the 10 voltages needed to each DM (9 low voltages for DM power and biasing the XA1 ASIC plus the HV bias for the CZT). The BCDH provides the control for the 8 DMs. It also collects the housekeeping and provides the communications to the IP (both the transfer of the event and housekeeping data to the IP, and the commands to control the DMs and Block from the IP).

The BCDH is basically a data concentrator, so that the IP need not communicate directly with the 128 DMs in the Detector Array. It receives photon and calibration event data from all 8 DMs and multiplexes this data into a single serial data stream that is transmitted to the IP. In the process, the BCDH embeds a DM identification in the event data so that the IP can determine which pixel produced the event from within the entire Detector Array. It also embeds timestamps so that the event data can be parsed and analyzed with respect to time. The BCDH is able to handle the event rate from a bright burst plus the background, which for BAT is about 12,000 events/sec/Block. This event rate represents the brightest burst BAT should see within a one year period.

The BCDH also receives and de-multiplexes commands from the IP, and transmits them to the specified DMs. These commands include: setting the bias voltages and currents for the DMs, setting the trigger threshold, enabling and setting the HV for the DMs, setting the operating modes (standby, normal, raw, or calibration) of the DMs, enabling individual channels within each DM, requesting and setting the housekeeping status cycle time intervals, resetting the DMs, and resetting the BCDH. The BCDH collects and reports back to the IP the housekeeping information associated with the DMs and the Block. These items include: all the bias voltages and currents for the DMs, the temperature of each DM, and the count rates of each DM.

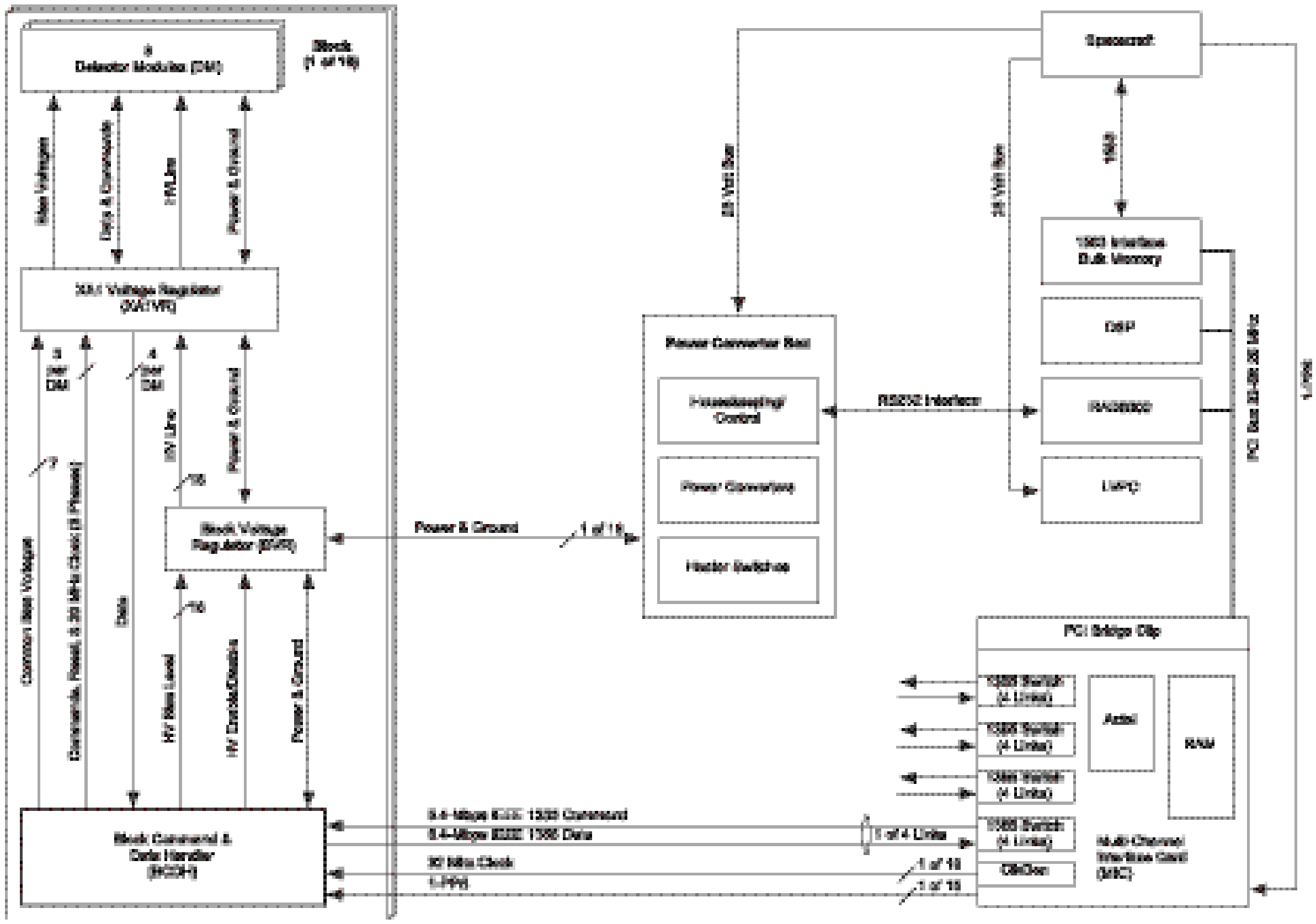

**Figure 5.** Context and block diagram for the Block, the Image Processor (IP), and the spacecraft. The Block Controller and Data Handler (left) controls the 8 DMs in the Block. The Power Converter Box (center) provides regulated power and the Image Processor Box (right) communicates events and commands with the Block and interfaces to the spacecraft (upper right).

### 2.1.2 The Coded Aperture Mask

It is not technologically possible to produce an image in the gamma-ray bandpass using traditional focusing optics; especially over a large FOV. Hence, the only way to formulate an image is to use the coded-aperture method. The BAT coded aperture is composed of ~52,000 lead tiles located 1 meter above the CZT detector plane. The Pb tiles are 5.00 mm square and 1.0 mm thick. The tiles are mounted on a low-mass, 5-cm thick composite honeycomb panel. The Mask panel is mounted above the Detector Plane by composite fiber struts (see Fig. 1). The pattern is completely random with a 50% open 50% closed filling factor. The commonly used Uniformly Redundant Array (URA) is not used for BAT because 1) the large FOV requirement means the aperture must be much larger than the detector plane, and 2) the detector plane is not uniform (i.e. there are gaps between the DMs). The Mask is 2.4 m by 1.2 m (with the corners cut off, it is 2.7 $m^2$), which yields a 100° by 60° FOV (half-coded). Figure 6 shows the shape of the 0, 50, and 100% coding contours and the encoded detector area as a function of the FOV. Figure 7 shows a picture of the BAT coded aperture mask.

### 2.1.3 The Fringe Shield

To reduce the event rate in the detector plane, a graded-Z Fringe Shield is located on the side walls between the Mask and the Detector Plane and under the Detector Plane. It reduces the isotropic cosmic diffuse flux and the anisotropic Earth albedo flux by ~95%. The graded-Z shield is composed of 4 layers of materials: Pb, Ta, Sn, and Cu; for a total mass of roughly 24 kg. In the side walls, the material layers are thicker towards the bottom nearest the Detector Plane, and get thinner towards the Mask. The side wall panels are mounted from the struts that hold the Mask.

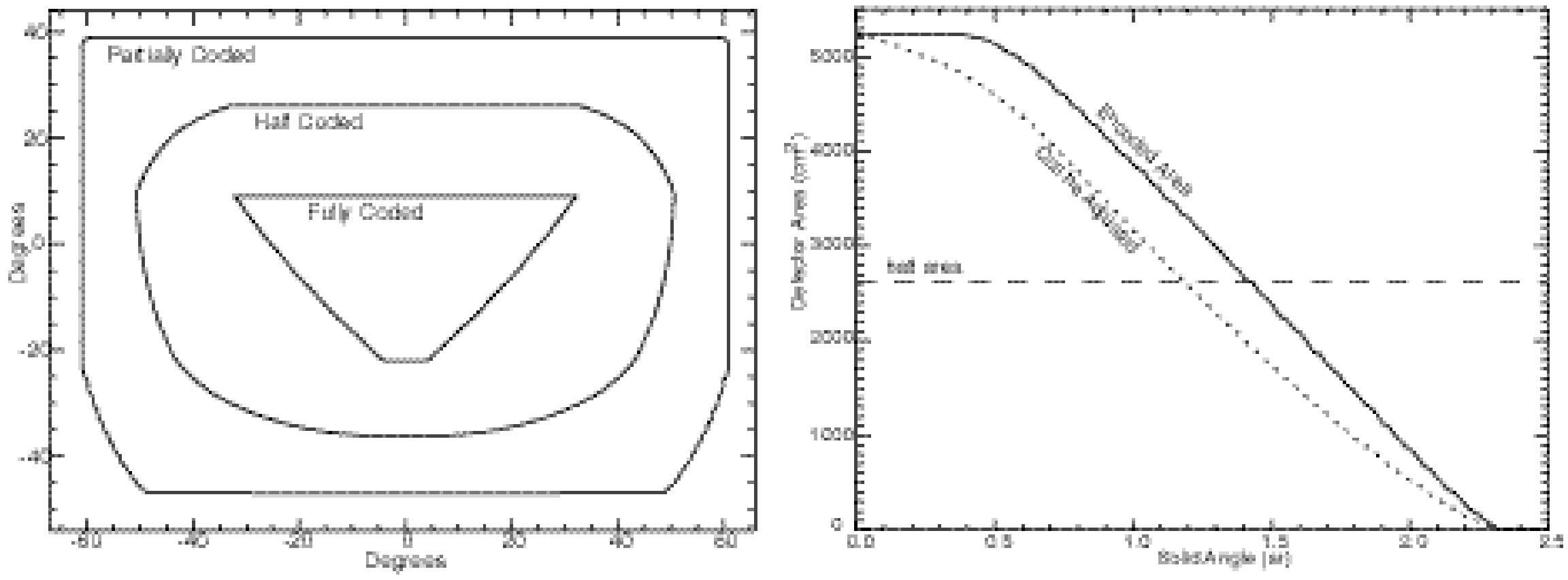

**Figure 6:** Plot of the BAT FOV with the 0%, 50%, and 100%-coding contours (left), and the geometric area of the Detector Plane as a function of solid angle (right).

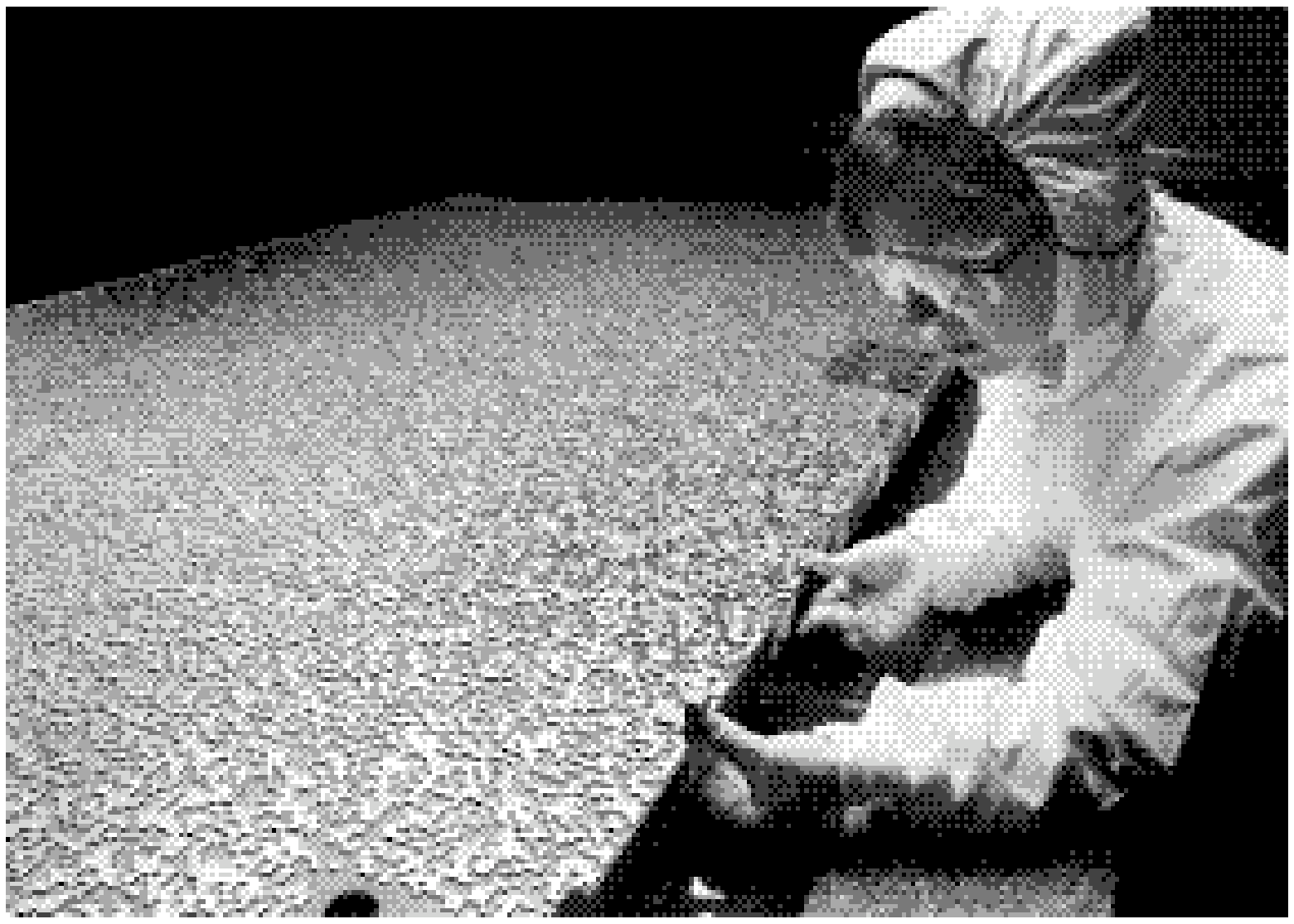

**Figure 7:** Image taken of the BAT's coded aperture mask.

### 2.1.4 The Thermal Control System

The CZT and XA1 front-end ASIC require the operating temperature to be controlled. The nominal operating temperature for the CZT is 20°C with a commandable range of 0-25°C. The temporal and spatial thermal gradients will be held to ±1°C. The close proximity of the XA1 to the CZT plus its internal power dissipation means that its operating temperature will be a few degrees higher than the CZT. This temperature difference is acceptable because it is constant with time and position across the DAP.

There are four parts to the DAP thermal control system: 1) the heaters with adjustable set points on each Block, 2) heat pipes embedded in the honeycomb plate under the Blocks, 3) heat pipes from the two side edges of the plate to the radiator, and 4) the thermal radiator mounted on the front side of the instrument. The radiator is 1.4 $m^2$ and is the only way the ~200 W dissipated in the DAP are removed to keep the temperature constant. The radiator is always on the anti-sun side of the spacecraft. The heat is brought to the radiator from the DAP by two Variable Capacity Loop Heat Pipes (VCLHP). The evaporation chambers of the Loop Heat Pipes are in contact with the ends of the 10 Constant Capacity Heat Pipes (CCHP) embedded in the detector array plate. The CCHP carry the heat dissipated in the Blocks by the detector electronics and the bias heaters to the VCLHPs. Even with the radiator mounted in the anti-sun side of the spacecraft, the thermal environment the radiator sees varies over an orbit due to different Earth viewing. There are 4 pairs of heater foils, each with its own commandable set-point, mounted on the side walls of the DMs in each Block. The radiator is sized to keep the DAP colder than the desired operating point under optimum conditions, and the VCLHP plus the heaters provide a variable conductance and heat source to match the radiator's variable ability to radiate the heat load to space as the Earth comes into and goes out of the radiator's 2À hemisphere.

### 2.1.5 The Image Processor

The Image Processor (IP) is the BAT instrument control processor. It does five major functions: 1) collects and scans the event stream looking for rate increases (i.e. bursts), 2) calculates the sky images when there is a rate increase and scans for new point sources, 3) determines if the newly detected burst is merits a slew request, 4) controls the instrument and gathers housekeeping information, 5) formulates the BAT-portion of the telemetry stream sent to the spacecraft control computer. These five functions are implemented in an architecture shown in Figure 5. Because of the requirement to generate the burst positions within 12 sec of the start of the burst, there are two processors within the IP. The RAD6000 is the main instrument processor. It handles the event processing, trigger searching, housekeeping, and command, control, and telemetry functions. The 21020 DSP does the FFT and back-projection image calculations (section 2.3.1.2). The 256 MB DRAM board provides all the memory for a 10-minute event-by-event ringbuffer, survey mode histograms, plus other science and engineering data products. The Multi-channel Interface Card (MIC) sends all the commands from the RAD6000 to the BCDH's and receives all the event and housekeeping data from the BCDHs. It uses the 1355 (Spacewire) implementation. All the data is DMA-ed into the ringbuffer in the DRAM card, where is it retrieved as needed by the RAD6000 for the trigger scanning and other functions. The 1553 board provide the interface between the BAT IP and the *Swift* spacecraft control computer.

## 2.2 Telescope Performance

Figure 8 shows the combined spectrum of the 32K detectors from the entire array for Co57. The FWHM of the 122 keV line is 6.2 keV. This resolution includes the intrinsic energy resolution of the CZT material and electronics and errors due to uncertainties in the gains and offsets when combining the individual spectra. The BAT background event rate has been calculated through computer simulation to be on average 17 Kevts/sec, with orbital variations of a factor of two around this value. This yields a GRB sensitivity of ~$10^{-8}$ erg $s^{-1}$ $cm^{-2}$. The BAT burst sensitivity will be five times better than BATSE. And as described in section 2.3.2, the hard x-ray survey will be 10-30 times more sensitive than the HEAO A-4 survey[8]. The combination of the 4 mm square CZT pieces, plus the 5 mm square Mask cells and the 1-m Detector-to-Mask separation yields an instrumental PSF of 17 arcmin FWHM. For an 8σ burst threshold, this yields a conservative 4 arcmin centroiding capability for bursts and steady-state sources -- stronger sources will be located to better than 4 arcmin.

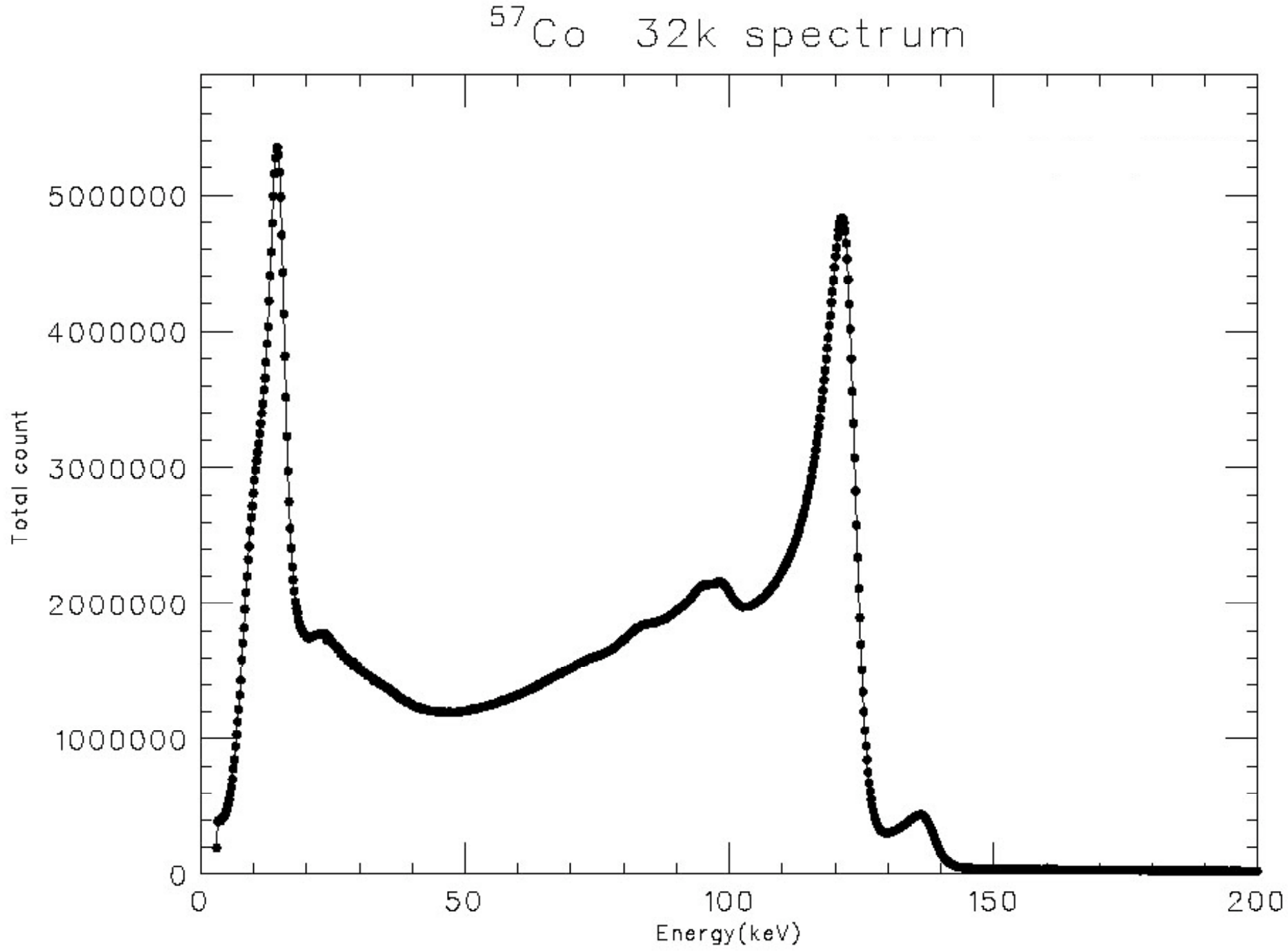

**Figure 8:** Composite spectrum of 32K CZT detectors (the entire Array) for Co57. The 14, 122, and 136 keV lines are clearly seen. The composite FWHM of the 122 keV lines is 7 keV.

## 2.3 Instrument Operations

The BAT instrument has two basic modes of operation: 1) scan-survey mode, and 2) burst mode. These two modes reflect the two major types of data that BAT produces: hard x-ray survey data and burst positions. Most of BAT's time is spent waiting for a burst to occur in its FOV. It accumulates events in the detector plane looking for increases in the count rate over a range of time scales. This scanning for rate increases is the trigger algorithm and is described in section 2.3.1.1. When the trigger algorithm is satisfied, it goes into burst mode (described in section 2.3.1). When not in burst mode and while scanning for a trigger, the instrument is accumulating spectra in each of its 32,786 detector elements every 5 minutes. These 32K spectra are recorded and become part of the survey data (section 2.3.2). During the scan-survey mode each block periodically goes into calibration mode. Within the BAT flight software is the Figure Of Merit calculation that decides if the current burst trigger is worth performing a spacecraft slew maneuver (section 2.3.3).

### 2.3.1 Burst Detection

Producing GRB locations within BAT is composed of two processes: 1) the detection of the onset of a burst by looking for increases in the event rate across the detector plan, and 2) the formation of an image of the sky using the events detected during the time interval at the beginning of the burst.

#### 2.3.1.1 Burst Trigger Algorithm

The burst trigger algorithm looks for excesses in the detector count rate above those expected from background and constant sources. The two main obstacles to GRB detection are the variation in background and the heterogeneity of GRB time profiles. In Low Earth Orbit, detector background rates can vary by more than a factor of two during a 90-minute orbit. The durations of GRBs range from milliseconds to minutes, during which they may have anywhere from one to several dozen peaks in the emission. Therefore, the triggering system must be able to extrapolate the background and compare it to the measured detector count rate over a variety of timescales and in several energy bands.

The trigger algorithms used in BAT are based on those developed for the HETE-2 GRB observatory. The algorithm continuously applies a large number of criteria which specify the pre-burst background intervals (typically 0-100 sec), the order of the extrapolation of the background rate (constant, linear, and parabolic with time), the duration of the burst emission test interval (4 msec to 32 sec), the region of the detector plane illuminated, and the energy range (typically 4 different bandpasses). A second burst detection method is also implemented. Every 64 sec the detector array count rate map is processed through the FFT imaging algorithm (on the DSP) and scanned for point sources. All sources found are compared against an on-board catalog. Any new sources or any known source with an intensity above a commandable level will constitute a "new source" and initiate the burst response procedure.

For example, one trigger criterion may specify that the count rate of the detectors on the left side of the detector plane in the 100-150 keV band over a 128 msec interval should be compared to the linear extrapolation of the background rate over a baseline interval from 5 to 3 seconds earlier. This searches for short, hard GRBs near the right edge of the field of view, even in the presence of a rising or falling background, but it is relatively insensitive to long, soft GRBs. The BAT processor will continuously track hundreds of these criteria sets simultaneously. The table of criteria can be adjusted after launch to balance sensitivity against the number of false triggers, and to concentrate on specific subclasses of GRBs as they are discovered.

#### 2.3.1.2 Burst Imaging and Location Process

Once the trigger algorithm detects a count-rate excess in the detector, the data are analyzed to discover if this is due to a GRB. The RAD6000 extracts source and background data based on the energy range and time intervals flagged by the trigger. For each detector it compares the count rate to the background, yielding a 32768-value detector map of count-rate excesses measuring the instrumental response to only the GRB transient source, i.e. the background subtraction removed the signal from the steady-state sources. This background-subtracted detector map is then sent to the 21020 DSP for further processing.

The DSP uses an FFT-based algorithm to correlate the pattern of detector counts with the coded aperture mask pattern. This requires about 12 seconds for the DSP to generate a 1024x512 pixel image showing the locations of transient sources. If a peak in this sky map is found, its position can be determined to within a single 14-arcmin sky pixel. A lack of an image peak may indicate insufficient flux to detect a burst, or it may indicate a false trigger, possibly produced by rapid variations in the particle background. A burst that is initially too weak to image may brighten considerably in the following seconds, and if so the trigger will once again request that imaging be attempted. The ability to use imaging to eliminate false triggers is a primary advantage of the BAT, and allows us to set the trigger thresholds to a sensitivity that would be intolerable if there were no other method of confirmation.

Once the approximate source location is known, the DSP executes a back-projection algorithm which produces an image of that region with arbitrarily small pixel size (typically 1 arcmin), giving a peak with the intrinsic 17 arcmin FWHM PSF of the instrument. Centroiding this peak gives the source location at the statistical and systematic limits of the instrumental accuracy (1-4 arcmin depending on the intensity of the burst). Figure 9 shows an image of the letters "BAT".

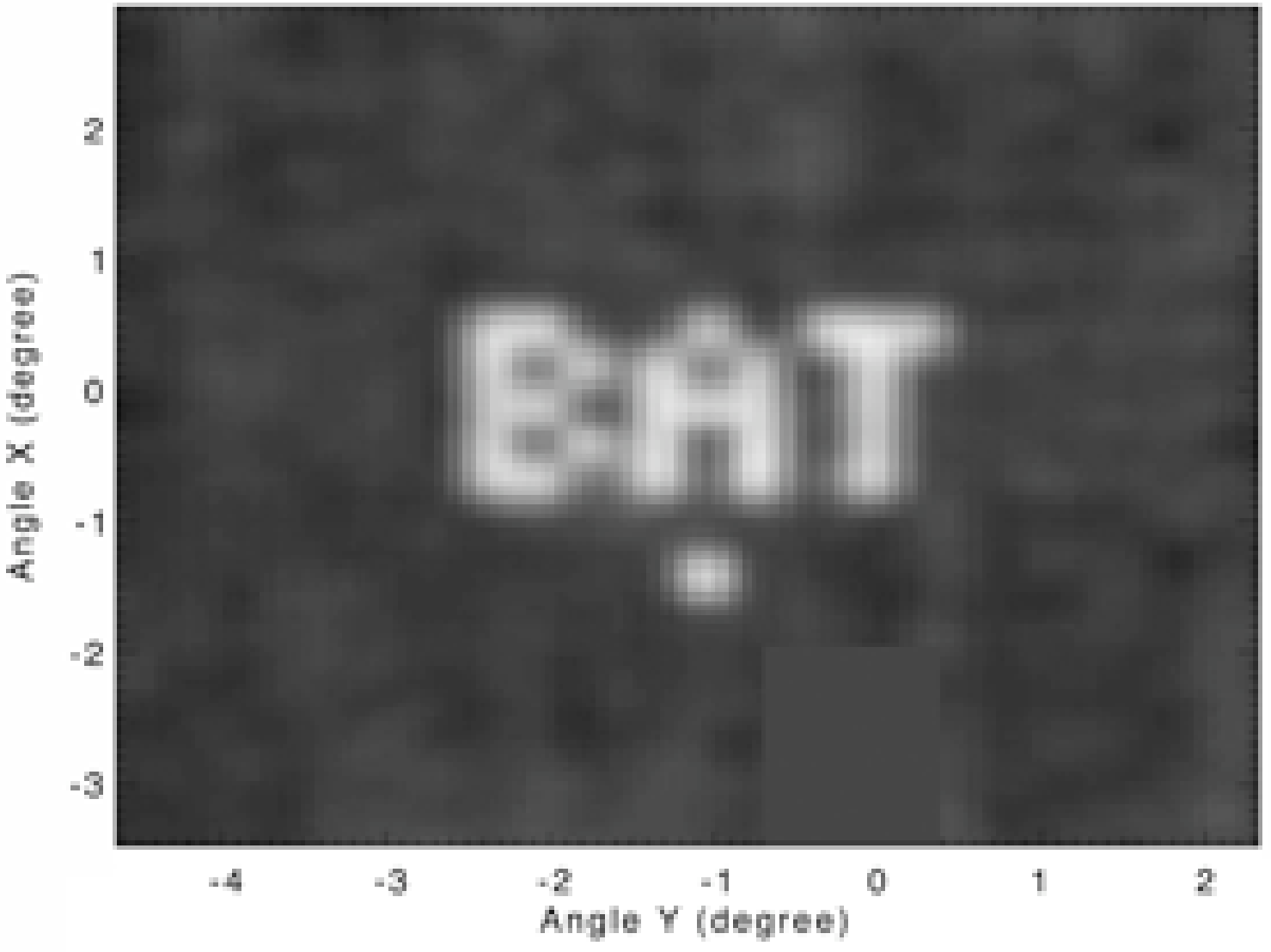

**Figure 9:** Image of the letters "BAT", demonstrating the burst imaging capabilities of the instrument.

#### 2.3.1.3 BAT Burst Data Products

The key aspect of the BAT instrument is to produce burst information quickly and get it distributed to the spacecraft and to ground-based observatories as soon as possible. Table 3 lists the burst-related data products, and how soon they become available. The burst trigger starts the time-available delay clock listed in the table. The Trigger Alert message alerts ground-based personnel that the BAT trigger algorithm has been satisfied. It contains the UT timestamp of the trigger and which time, energy, and geometry trigger was satisfied. This alert message is immediately transferred to the spacecraft processor, and then the TDRSS transponder is powered up. Meanwhile the BAT continues to process the event data to produce a location in right ascension and declination, which is also transferred to the spacecraft and transmitted to the ground through TDRSS. BAT will produce burst location error circles with a 4 arcmin diameter and have them available on the ground within 20 sec of the start of the burst. The results of the FOM decision and the spacecraft decision are also transmitted. After 130 sec, ~30 sec of pre-trigger and ~120 sec of post-trigger light curve information in 4 energy bandpasses are transmitted. The XRT[6] and UVOT[7] instruments also produce position, spectra, and images of the burst afterglow, which also go through TDRSS to the ground and through GCN (see section 3).

**Table 3:** BAT burst-related data products and their time delays after the burst trigger.

| DATA PRODUCT | TIME AVAILABLE |
|---|---|
| Trigger Alert | 5 sec |
| Burst Position | 12 sec |
| FOM Will/Wont Observe | 14 sec |
| S/C Will/Wont Slew | 14 sec |
| Burst Light Curve | 130 sec |
| Burst afterglow lightcurve | ~ 20 min |
| Burst event-by-event data | 2-4 hrs |

### 2.3.2 Hard X-ray Survey

While detecting and chasing bursts, BAT will perform an all-sky hard x-ray survey and be a monitor for hard x-ray transients. BAT accumulates detector plane maps (i.e. an energy spectrum in each of the 32K detectors) every 5 minutes. These 5-min spectra maps are continuously included in the normal spacecraft telemetry stream. On the ground, sky images are produced by FFT convolution with the mask pattern from the survey data using an iterative clean algorithm. These sky images are searched with a peak-detection algorithm to find and localize sources. For each detected source, a back-projection algorithm produces a background-subtracted counts spectrum and a refined position. Fits will be performed to the counts spectrum using a detailed instrument response matrix including the off-diagonal response to determine the source photon spectrum.

With the BAT instrument parameters, the statistical sensitivity is about 0.2mCrab in the 15-100 keV band for 3 years. Deep all-sky surveys are dominated by systematic errors, however. For regions in the Galactic Plane (l<45°) where there are several strong sources always in the BAT FOV, the limiting sensitivity will be ~2 mCrab,, which is reached in about a day of accumulation. For l>45°, the limiting sensitivity will be ~0.6 mCrab. This is 10x and 30x, respectively, more sensitive than the previous best all-sky survey by HEAO A-4 of ~17 mCrab[8]. For known sources >2 mCrab, BAT will be able to measure a 3σ flux in less than a day, thus providing a detailed light curve of hundreds of sources. Figure 10 shows the time required to achieve a 5σ detection sensitivity for sources with different fluxes.

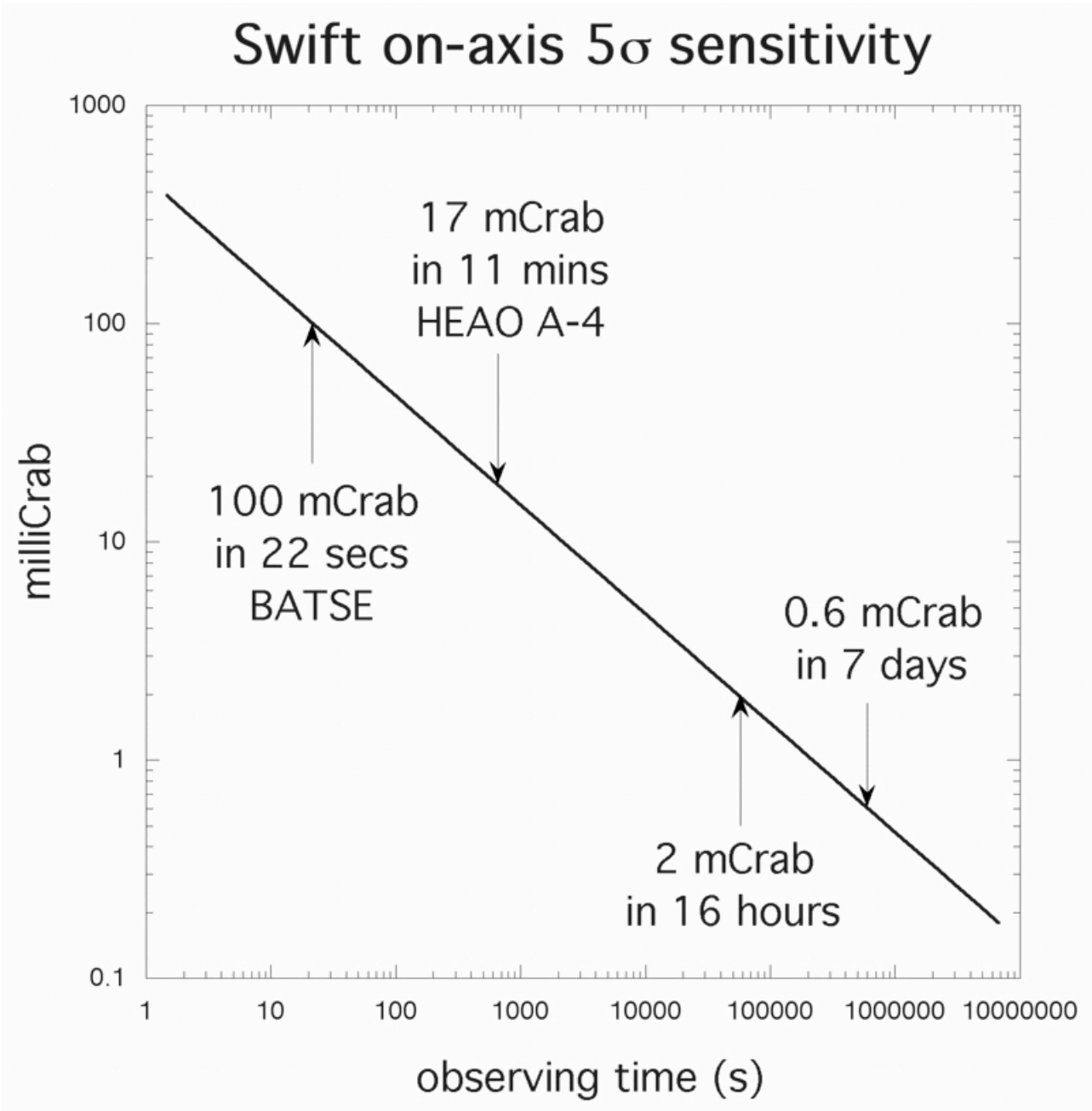

**Figure 10:** Time required to achieve a 5σ detection sensitivity for sources with different fluxes, assuming a Crab spectrum.

### 2.3.2.1 Hard X-ray Transients

For on-board transient detection, 1-minute detector plane count-rate maps are accumulated in 4 energy bandpasses. Each 1-minute map is passed through the FFT image construction process (see section 2.3.1.2 for the details on the FFT procedure). The sources found in these images are compared against an on-board catalog of sources. Sources not listed in the catalog or sources showing large variability are deemed transients. This process is also carried out on longer time scales of 5 minutes and for the total time between spacecraft slews to a new target in the observing table (typically 10-40 minutes). We expect this process also to detect a subclass of long, smooth GRBs that will not be detected by the burst rate-trigger algorithm. All hard x-ray transients will be distributed to the world community through the GCN system, just like the bursts.

### 2.3.3 Figure of Merit Algorithm

The Figure of Merit (FOM) algorithm is part of the spacecraft's autonomy that decides if the burst just detected by BAT is worth requesting a slew maneuver by the spacecraft. While not strictly part of the BAT instrument -- it is more of a mission-wide function -- it does reside within the BAT Flight Software on the BAT Image Processor, and therefore will be described here. With each new burst detected and localized by BAT, the FOM determines through a series of criteria if this new burst is worth interrupting the current follow-up observation by the XRT and UVOT. If the new burst has more "merit", a slew request is sent to the spacecraft, which then checks the constraints and informs the FOM if the slew will be done. The FOM also processes uploaded TOO positions exactly the same as BAT-derived burst positions. The criteria for determining the merit of the newly detected burst and for the prior burst under follow-up observation by the NFI is based on 1) the brightness of the bursts, 2) the type of trigger detection, and 3) the observability of the position (sun/moon constraints, etc). The FOM is implemented entirely in software and can be changed (the current criteria can have their parameters adjusted and new criteria can be added). Figure 11 shows a model for the effective area of BAT. This is based on the measured effective area at 4 points, using a Ba-133 source. The mask-weighted count rates are added so as to include counts from the 31, 35, 53, and 80 keV lines. The fraction of the total count rate coming from each line was determined from the incident flux at each energy, the quantum efficiency of the CZT detectors at each energy, the transmission of each line through all passive materials, and the fraction of counts not in the escape peaks.

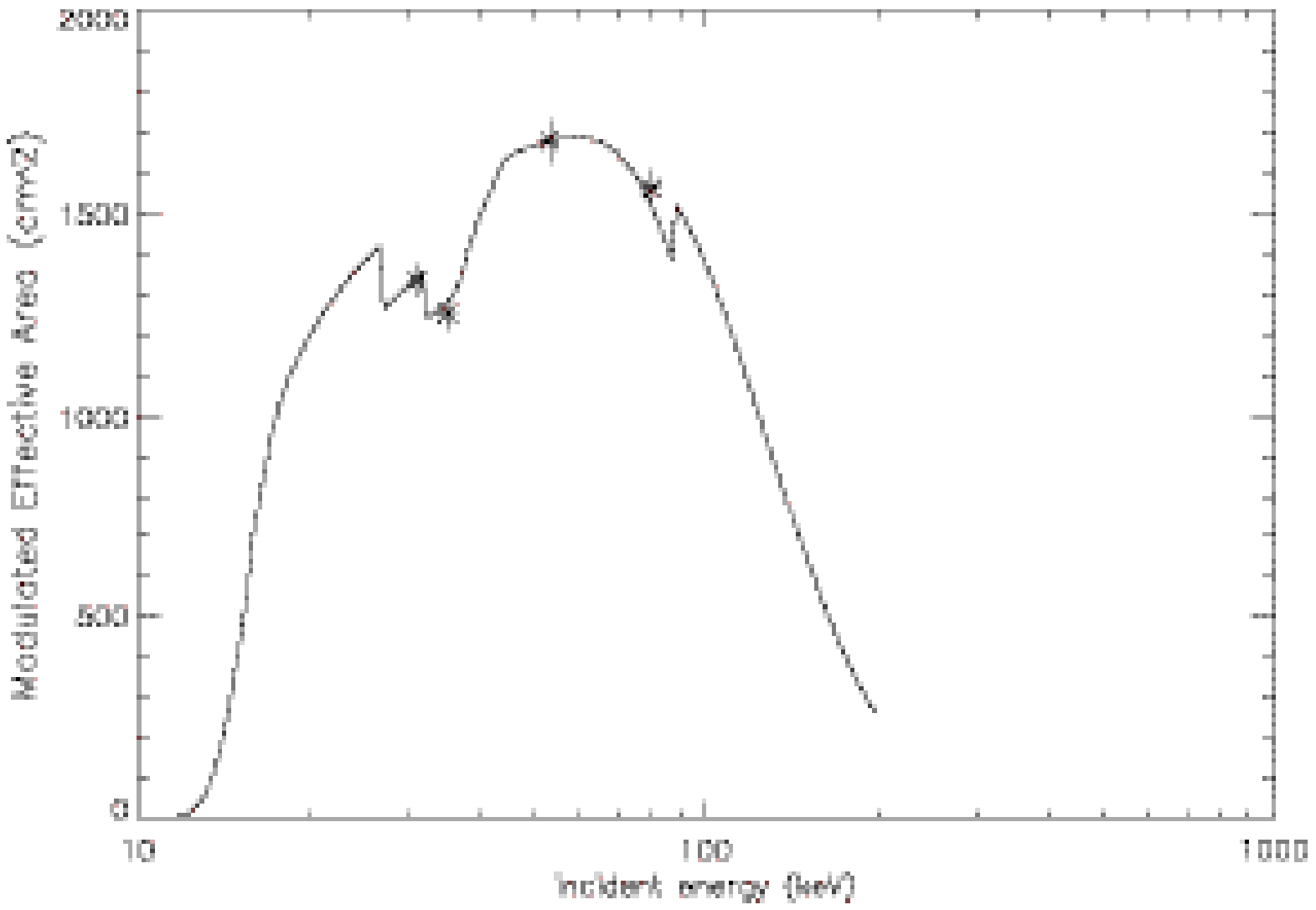

**Figure 11:** Model for the effective area of BAT.

## 2.4 BAT Instrument Status

As of September 2004, the BAT instrument has been successfully integrated into *Swift,* which was delivered to Florida in July. Launch is currently scheduled for November 2004.

# 3. GROUND SYSTEM AND DATA ANALYSIS

## 3.1 Mission and Instrument Control

The Mission Operations Center (MOC) is located at Pennsylvania State University. The MOC is responsible for the day-to-day health and safety monitoring and command control of the *Swift* mission -- both the spacecraft and the 3 instruments. This includes the general mission-science planning, the daily observing table up-loads, etc. Routine health and safety issues will be detected and handled by semi-automated operations software. And the more serious problems will generate alarms and pager messages will be issued to the appropriate personnel to respond to the problem. It is also responsible for the Level-0 data processing (time ordering the telemetry data stream and removing duplicate transmissions).

## 3.2 Ground Processing

There are two parts to the ground processing for the BAT instrument: 1) the real-time TDRSS burst messages, and 2) the production pipe-line processing for the daily telemetry dumps from the Malindi ground station.

### 3.2.1 The Real-Time Burst Message Processing

The real-time generated position, lightcurve, spectral, and image information generated by BAT (and XRT, and UVOT) is transmitted to the ground via the Multiple Access (MA) system within TDRSS. Once on the ground they are transferred to the GCN system[9,10], and distributed to any instrument, person, or institution wishing the information (to make follow-up observations, or for general awareness about the current state of GRB activity).

Figure 12 shows the flow of telemetry data from *Swift* to the ground 1) for the TDRSS real-time Message Alert system, 2) for the normal (full data set) telemetry down-links, and 3) for the command uplink. The path of interest for the GCN operation is the TDRSS MA real-time Message link. The TDRSS link path is from the *Swift* spacecraft TDRSS transponder to the closest of the 3 in-orbit TDRSS satellites, which then gets retransmitted to the White Sands ground station in New Mexico. From there the messages get passed to the GCN system[9,10] at Goddard through an Internet socket connection. The MOC (PSU) also has a parallel socket connection from White Sands. The total time from the start of the burst (the BAT trigger) to the time the BAT burst positions are available to the world community is 20 seconds – the other burst-related TDRSS data products will take slight longer due to the extra time delays in their generation on board the spacecraft.

### 3.2.2 Quick-Look and Production Processing

There are two main stages in the production pipeline data processing: 1) the quick-look processing which happens within a few hours after receipt of the data from Malindi, and 2) the final production processing which occurs about a week later. Both stages produce high-level data products for use by the instrument teams and the public. While both incorporate BAT instrument calibration response functions, the quick-look processing does not make use of all the cleaning, correction, and calibration effort that is ultimately used in the production processing. Rather, the quick-look is meant to get most of the value-added input into the data products, but clearly not all possible added value that is permitted by the extra week of processing and human intervention. With 7-10 telemetry contacts per day and the 3 hours for the quick-look processing, the typical time delays after a given burst for the availability of these data products will be in the range of ~4 to 24 hours. This makes the lightcurve, spectra, and images available to the public while ground-based follow-up activities are in progress or in the planning stages. These high-level data products will be available via web pages and ftp from the HEASARC, UK data center, and the Italian data center. They will be in standard FITS format, and the standard analysis tools can be used (e.g. ftools).

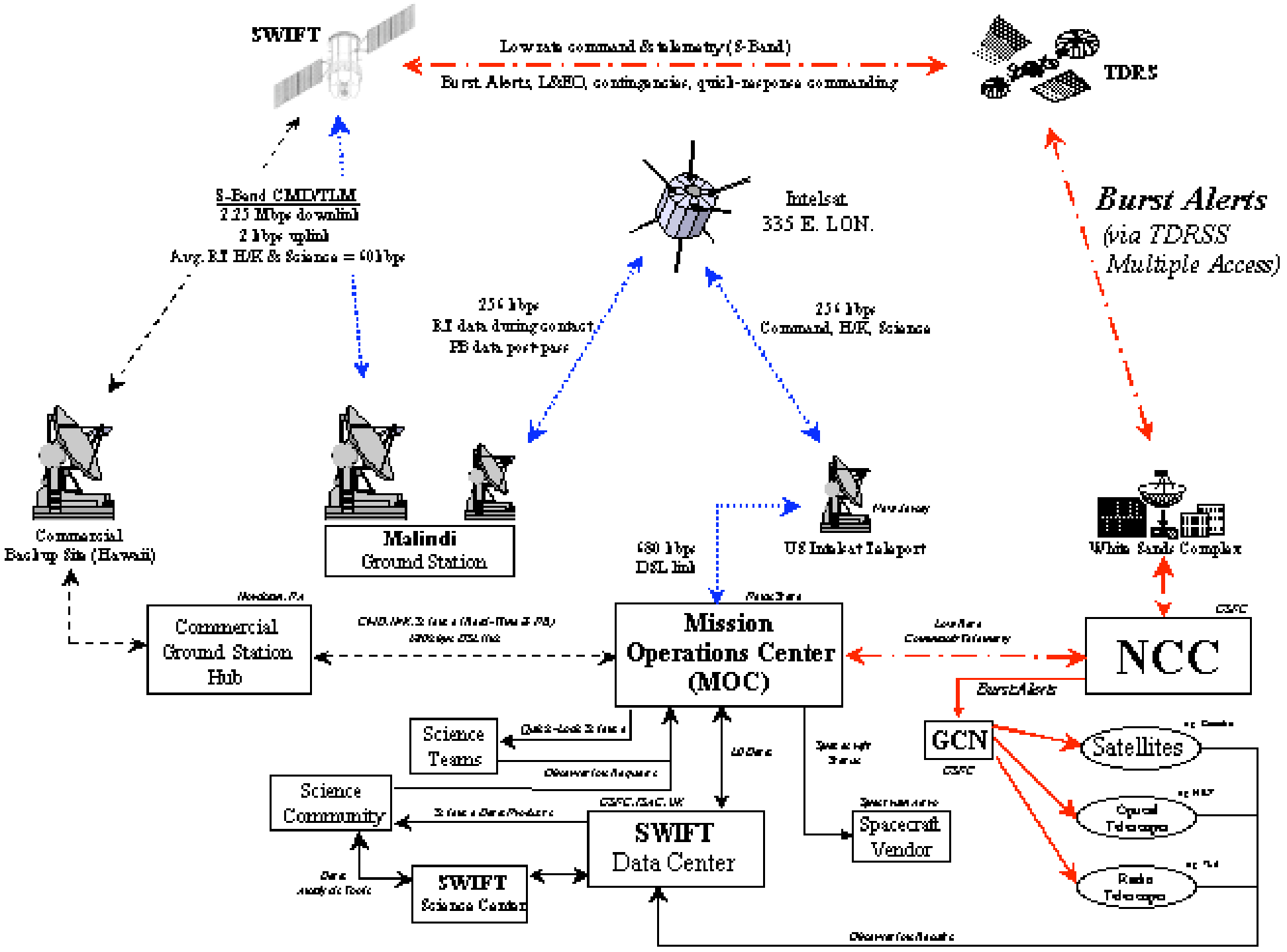

**Figure 12:** Communications links between the *Swift* spacecraft and the ground for the regular Malindi command/telemetry and for the TDRSS Messages link (and to the GCN).

## ACKNOWLEDGEMENTS

The core of the BAT instrument design team consists of the following engineers, technicians, and scientists: Robert Baker, Louis Barbier, Scott Barthelmy, Mike Blau, Frank Birsa, Dale Brigham, Jim Caldwell, Mike Choi, Jay Cummings, Mitch Davis, Norman Dobson, Jeff Dumontheir, Ed Fenimore, Greg Frazier, Oscar Gonzalez, Steve Graham, Ken Harris, Larry Hilliard, Jerry Hengemihle, Frank Kirchman, Hans Krimm, Rick Mills, Burt Nahory, Jim Odom, John Ong, David Palmer, Brad Parker, Ann Parsons, Kiran Patel, Traci Pluchak, Dave Robinson, Mark Secunda, Beverly Settles, Frank Shaffer, Dave Sheppard, Sandy Shuman, Ken Simms, Miles Smith, Steve Snodgrass, Dave Sohl, Carl Stahle, Jack Tueller, Danielle Vigneau, Charles Wildermann, and George Winkert. The authors would like to thank J. K. Cannizzo and J. D. Myers in the preparation of this manusscript.